\documentclass{IEEEtran}
\IEEEoverridecommandlockouts
\usepackage{array}
\usepackage{subfigure}
\usepackage{float}
\usepackage{amsfonts,amssymb,amsmath,calc,graphicx}
\usepackage{psfrag,epsfig,verbatim,moreverb,mathrsfs}
\usepackage{bm}
\usepackage{color}
\usepackage{epstopdf}

\usepackage{cases}

\def\T{\mathcal{T}}
\def\1{\textbf{1}}
\newtheorem{theorem}{Theorem}
\newtheorem{corollary}{Corollary}
\newtheorem{definition}{Definition}

\begin{document}

\title{Log-Convexity of Rate Region in 802.11e WLANs}
\author{Douglas J. Leith, Vijay G. Subramanian and Ken R. Duffy\\Hamilton Institute, National University of Ireland Maynooth\thanks{Work supported by Science Foundation Ireland grant 07/IN.1/I901. }}
\maketitle

\begin{abstract}
In this paper we establish the log-convexity of the rate region in 802.11 WLANs.   This generalises previous results for Aloha networks and has immediate implications for optimisation based approaches to the analysis and design of 802.11 wireless networks.
\end{abstract}

\section{Introduction}

\begin{figure}
\centering
\includegraphics[width=0.7\columnwidth]{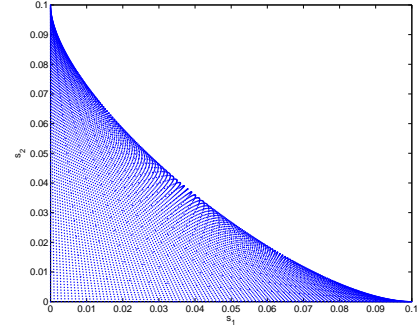}
\caption{Illustrating non-convexity of 802.11 rate region.  Plot shows throughput normalised by PHY rate for $n=2$ stations and $\sigma/T_c=1/10$ and $T_s=T_c$ (i.e. for packet sizes where the packet transmission duration is 10 times larger than the PHY idle slot duration).}\label{fig:rateregion}
\end{figure}

\begin{figure}
\centering
\includegraphics[width=0.7\columnwidth]{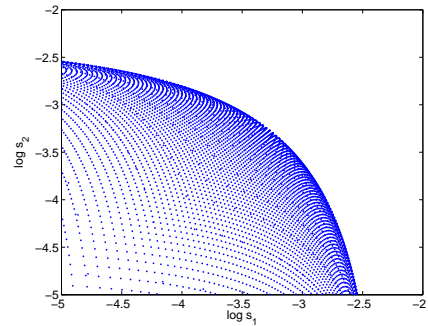}
\caption{Log rate region corresponding to data shown in Figure \ref{fig:rateregion}.}\label{fig:lograteregion}
\end{figure}

In this paper we consider the log-convexity of the rate region in 802.11 WLANs.    The rate region is defined as the set of achievable throughputs and we begin by noting that the 802.11 rate region is well known to be non-convex.  This is illustrated, for example, in Figure \ref{fig:rateregion} for a simple two-station WLAN (where $\sigma, T_c, T_s$ are described in Section~\ref{sec:nm}).  The shaded region indicates the set of achievable rate pairs ($s_1$, $s_2$) where $s_i$ is the throughput of station $i$, $i\in\{1,2\}$.  It can be seen from this figure that the maximum throughput achievable by the network when only a single station transmits (the extreme point along the x- or y-axes) is greater than that when both stations are active (e.g. the extreme point along the $y=x$ line).    This non-convex behaviour occurs because in 802.11 there is a positive probability of colliding transmissions when multiple stations are active, leading to lost transmission opportunities.   In Figure \ref{fig:lograteregion}  the same data is shown but now replotted as the log rate region, i.e. the set of 
pairs ($\log s_1$, $\log s_2$).   Evidently, the log rate region is convex.   Our main result in this paper is to establish that this behaviour is true in general, not just in this particular example.  That is, although the 802.11 rate region is non-convex, it is nevertheless log-convex.    
The implications of this for optimisation-based approaches to the design and analysis of fair throughput allocation schemes are discussed after the result.

In a WLAN context, rate region properties have mainly been studied for Aloha networks.
The log-convexity of the Aloha rate region in general mesh network settings  has been established by several authors  \cite{wang04,tassiulas04,calderbank06,sasha06,wang06} in the context of utility optimisation.    All of these results make the standard Aloha assumption of equal idle and busy slot durations, whereas in 802.11 WLANs highly unequal slot durations are the norm e.g. it is not uncommon to have busy slot durations that are 100 times larger than the PHY idle slot duration.    This is key to improving throughput efficiency but also fundamentally alters other throughput properties since the mean MAC slot duration and achieved rate are now strongly coupled.  We note that a number of recent papers have considered  algorithms that seek to achieve certain fair solutions (proportionally fair, max-min fair)  in 802.11 networks, e.g see \cite{siris06} and references therein.   For the WLAN scenario in this paper we show how existence and uniqueness of fair solutions follows from log-convexity.

\section{Network Model}\label{sec:nm}

The 802.11e standard extends and subsumes the standard 802.11 DCF (Distributed Coordinated Function) contention mechanism by allowing the adjustment of MAC
parameters that were previously fixed.   With 802.11, on detecting the wireless
medium to be idle for a period $DIFS$, each station initializes a counter to a
random number selected uniformly in the set \{0, ...,CW-1\} where CW is the 
contention window.  Time is slotted and this counter is decremented once for each slot that
the medium is idle.  An important feature is that the countdown halts when the
medium becomes busy and only resumes after the medium is idle again for a period
$DIFS$.  On the counter reaching zero, the station transmits a packet.  If a
collision occurs (two or more stations transmit simultaneously), CW is set to $\min(2\times CW, CW_{max})$  and the process repeated. On a successful transmission, CW is reset to the value
$CW_{min}$ and a new countdown starts for the next packet. Again, each packet transmission in this phase includes the time spent waiting for an acknowledgement from the receiver. 
The 802.11e MAC enables
the values of $DIFS$ (called $AIFS$ in 802.11e), $CW_{min}$ and $CW_{max}$ to be set on a per
class basis for each station.   Throughout this paper we restrict attention to situations where $AIFS$ has the legacy value $DIFS$.  In addition, 802.11e adds a TXOP mechanism that specifies the duration during which a station can keep transmitting without releasing the channel once it wins a transmission opportunity. In order not to release the channel, a SIFS interval is inserted between each packet-ACK pair. A successful transmission round consists of multiple packets
and ACKs. By adjusting this time, the number of packets that may be transmitted by a
station at each transmission opportunity can be controlled.  A salient feature of
the TXOP operation is that, if a large TXOP is assigned and there are not enough
packets to be transmitted, the TXOP period is ended immediately to avoid wasting
bandwidth.

We consider an 802.11e WLAN with $n$ stations.   As described in \cite{David_TON_2007,clifford06}, we divide time into MAC slots, where each MAC slot may consist either of a PHY idle slot, a successful transmission or a colliding transmission (where more than one station attempts to transmit simultaneously).  Let $\tau_i$ denote the probability that station $i$ attempts a transmission.  The mean throughput of station $i$ is then shown in \cite{David_TON_2007} to be
\begin{align}
s_i (\T) = \frac{\tau_i \prod_{k\in  N\setminus\{i\}} (1-\tau_k)L_i}{\sigma P_{idle} + T_sP_{succ}+T_c(1-P_{idle}-P_{succ})} \label{eq:tput_exp}
\end{align}
where $P_{idle}=\prod_{k\in N} (1-\tau_k)$ and $P_{succ}=\sum_{i\in N}\tau_i \prod_{k\in  N\setminus\{i\}} (1-\tau_k)$, $\T=[\tau_1\ ...\ \tau_n]^T$, $L_i$ is the mean frame payload size at station $i$ in bits and $N=\{1,..,n\}$, $\sigma$ is the PHY idle slot duration, $T_s$ is the duration of a successful transmission (including time to transmit the data frame, receive the MAC ACK and wait for DIFS) and $T_c$ the duration of a collision. In this paper we prove useful analytical properties of the throughput expression (\ref{eq:tput_exp}).

It will prove useful to work in terms of the quantity $x_i=\tau_i/(1-\tau_i)$ rather than $\tau_i$.  With this transformation we have that $P_{idle}=1/\prod_{k\in N} (1+x_k)$ and $P_{succ}=\sum_{i\in N} x_i/\prod_{k\in N} (1+x_k)$ and so
\begin{align*}
s_i (\T) = \frac{x_iL_i/T_c}{\sigma/T_c-1  + (T_s/T_c-1)\sum_{i\in N} x_i+\prod_{k\in N} (1+x_k)}
\end{align*}


\begin{definition} \emph{Rate Region}. The rate region is the set $R(\bar{\tau})$ of achievable throughput vectors $S(\T)=[s_1\ ...\ s_n]^T$ as the vector $\T$ of attempt probabilities ranges over domain $D(\bar{\tau})=[0,\bar{\tau}_1]\times\cdots\times[0,\bar{\tau}_n]$, where $\bar{\tau}_i$ denotes the $i$'th element of vector $\bar{\tau}$ and $0\le \bar{\tau}_i\le 1$, $\forall i\in\{1,...,n\}$.    
\end{definition}

In this paper we assume that the value of $\tau_i$ can be freely selected in the interval $[0, \bar{\tau}_i]$.  This is a mild assumption.  For example, suppose  $CW_{max}$ is set equal to $CW_{min}$.  Then\footnote{Ignoring post backoff for simplicity} $\tau = 2q/CW_{min}$ where $q$ is the probability that there is a packet available for transmission when the station wins a transmission opportunity and so is related to the packet arrival rate.   When a station is saturated we have $q=1$.   We note that the value $q$ here is similar to the quantity in \cite{David_TON_2007} also referred to as $q$.   By adjusting $q$ (via the packet arrival process) and/or $CW_{min}$, it can be seen that the value of $\tau_i$ can be controlled as required.\newline

\begin{definition}\emph{Log-convexity}.  Recall that a set $C\in \mathbb{R}^n$ is convex if for any $s^1, s^2 \in C$ and $0\le \alpha \le 1$, there exists an $s^* \in C$ such that $s^*=\alpha s^1 + (1-\alpha) s^2$. A set $C$ is log-convex if the set $\log C := \{\log s: s\in C\}$ is convex.
\end{definition}

\section{Log-Convexity}

\subsection{Log-Convexity}
%
%

We begin in this section by assuming that $\bar{\tau}=\1$, where $\1$ denotes the all 1's vector.  This assumption is relaxed later on.  For convenience we set $a:=\sigma/T_c$ with $a\in [0,1]$ and $K:=T_s/T_c-1$ with $K\geq 0$. The throughput expression can now be written as
\begin{align}\label{eq:tput}
s_i(\T)=\frac{x_i L_i/T_c}{X(\T)}
\end{align}
where 
\begin{align}\label{eq:Xdef}
\begin{split}
X(\T) & :=a+K \sum_{i\in N} x_i + \prod_{i\in N} (1+x_i) -1 \\
& \;= a+(K+1) \sum_{i\in N} x_i + \sum_{k=2}^n \sum_{A\subseteq N: |A|=k} \prod_{j\in A} x_j.
\end{split}
\end{align} 

We know that the rate region $R(\1)$ may be non-convex, but ask whether it is log-convex.   Let $\log S(\T) = [\log s_1\ ...\ \log s_n]^T$. The rate region $R(\1)$ is log-convex if $\forall\; \T^1, \T^2 \in (0,1)^n$ and $\forall\; \alpha\in[0,1]$, $\exists \T^* \in (0,1)^n$ such that
\begin{align}\label{eq:sdef}
\alpha \log S(\T^1) + (1-\alpha) \log S(\T^2) = \log S(\T^*).
\end{align}
Rearranging terms we get for every $i=1,\dotsc,n$,
\begin{align}
\frac{x_i^*}{X(\T^*)}
& = \left( \frac{x_i^1}{X(\T^1)}\right)^\alpha\left(\frac{x_i^2}{X(\T^2)} \right)^{(1-\alpha)}, \text{ or}\notag \\
\label{eq:first} 
\frac{(x_i^1)^\alpha (x_i^2)^{(1-\alpha)}}{x_i^*} &= \frac{X(\T^1)^\alpha X(\T^2)^{(1-\alpha)}}{X(\T^*)}.
\end{align}
Note that here we restrict $\T$ to $(0,1)^n$ rather than $[0,1]^n$.  This involves no loss of generality since $S(\T)$ is a continuous function of $\T$.  Note that the $L_i/T_c$ term in (\ref{eq:tput}) cancels on both sides of (\ref{eq:sdef}) so the log-convexity result is independent of this term.

We proceed by postulating that $x^*$ is of the form
\begin{equation}\label{eq:soln}
x_i^*=\frac{(x_i^1)^\alpha (x_i^2)^{(1-\alpha)}}{\delta}
\end{equation}
as the right side of (\ref{eq:first}) does not depend on any particular $i$.
The log-convexity question is whether we can find $\delta>0$ satisfying
\begin{equation}\label{eq:delta}
\delta=\frac{X(\T^1)^\alpha X(\T^2)^{(1-\alpha)}}{X(\T^*)}
\end{equation}
Substituting from (\ref{eq:soln}) into (\ref{eq:delta}), then using the first expression in (\ref{eq:Xdef}), and defining $y_k=(x_k^1)^\alpha (x_k^2)^{(1-\alpha)}$, we will need to solve for a $\delta > 0$ such that
\begin{align}
\delta& =\frac{X(\T^1)^\alpha X(\T^2)^{(1-\alpha)}}{a+K \sum_{i\in N} \frac{y_i}{\delta} + \prod_{i\in N} \left(1+\frac{y_i}{\delta}\right) -1}, \text{ i.e.} \notag \\
\label{eq:soln2}
\begin{split}
&\delta\left(a+K \sum_{i\in N} \frac{y_i}{\delta} + \prod_{i\in N} \left(1+\frac{y_i}{\delta}\right) -1\right)  \\
&\qquad \qquad \qquad \qquad ={X(\T^1)^\alpha X(\T^2)^{(1-\alpha)}}.
\end{split}
\end{align}
Recalling H\"olders inequality for two non-negative vectors $u$ and $v$,
\begin{align*}
\left( \sum_k u_k\right)^\alpha \left(\sum_k v_k\right)^{(1-\alpha)}  \ge \sum_k u_k^\alpha v_k^{(1-\alpha)} \; \forall \alpha \in [0,1],
\end{align*}
we have using the second expression in (\ref{eq:Xdef}) that the right-hand side of (\ref{eq:soln2}) is positive and lower bounded by
\begin{align*}
a+K \sum_{i\in N} y_i + \prod_{i\in N} \left(1+y_i\right) -1.
\end{align*}
Choosing $\delta=1$ it can be seen that this lower bound lies within the range of the left-hand side of  (\ref{eq:soln2}). Considering the left-hand side of (\ref{eq:soln2}) in more detail, its second derivative is given by
\begin{align*}
\frac{1}{\delta^3} \sum_{i, j\in N: j\neq i} y_i y_j \prod_{k\in N: k\neq i, j} \Big(1+\frac{y_k}{\delta}\Big)  
\end{align*}
where product over an empty set is defined to be $1$. Since the second-derivative is positive for $\delta \geq 0$, it implies the (strict) convexity of the left-hand side of (\ref{eq:soln2}).
This quantity is unbounded and has range that includes $[a+K \sum_{i\in N} y_i + \prod_{i\in N} \left(1+y_i\right) -1, \infty)$. It follows that there exists a positive $\delta$ satisfying (\ref{eq:soln2}), as required. Indeed, in general there may exist two values of $\delta$ solving (\ref{eq:soln2}). To see this observe that the left -hand side is unbounded both as $\delta\rightarrow 0$ and as $\delta\rightarrow\infty$. The first-derivative is negative as $\delta\rightarrow 0$ and positive as $\delta\rightarrow \infty$, so we have a turning point $\delta^*$, which due to the convexity of the function is unique. This turning point partitions the real line and two solutions to (\ref{eq:soln2}) then exist, one lying in $(0,\delta^*)$ and the other in $(\delta^*,\infty)$. Additionally, this argument also says that there exists at least one solution of (\ref{eq:soln2}) where $\delta \geq 1$.

We have therefore established the following theorem.
\begin{theorem}\label{thm:one}
The rate region $R(\1)$ is log-convex.
\end{theorem}
  
\subsection{Constraints on $\tau$}

We can extend the foregoing analysis to situations where the station attempt probability is constrained, i.e. the vector $\T$ of attempt probabilities ranges over $D(\bar{\tau})=[0,\bar{\tau}_1]\times\cdots\times[0,\bar{\tau}_n]$, where $0\le \bar{\tau}_i\le 1$, $\forall i\in\{1,...,n\}$. Note that an upper bound on $\tau_i$ of $\bar{\tau}_i$ results in an upper bound $\bar{x}_i=\bar{\tau}_i/(1-\bar{\tau}_i)$ on $x_i$. Therefore if $\T^1, \T^2\in \bar{\tau}$, then $x^1, x^2 \in D(\bar{x})=[0,\bar{\tau}_1/(1-\bar{\tau})_1]\times\cdots\times[0,\bar{\tau}_n/(1-\bar{\tau})_1]$ and for every $\alpha \in [0,1]$ we also have $y\in D(\bar{x})$. From the proof of Theorem~\ref{thm:one} we know that there exists at least one $\delta \geq 1$ that solves (\ref{eq:soln2}). Using that solution we find that $x^*=y/\delta \leq y$ so that
$x^*\in D(\bar{x})$. Note that we can have different values of $\bar{\tau}_i$ for every $i$. Therefore we have the following corollary to Theorem~\ref{thm:one}.
\begin{corollary}
The rate region $R(\bar{\tau})$ is log-convex for every $\bar{\tau}\in [0,1]^n$.
\end{corollary} 



\section{Discussion}

These log-convexity results allow us to immediately apply powerful optimisation results to the analysis and design of fair throughput allocations for 802.11 WLANs.  First, using  \cite[Theorem~1]{boudec07}, the existence of a max-min fair solution immediately follows. We also have that any optimisation of the form
\begin{align*}
\max_S f(S) \text{ s.t. } \; S \in R(\bar{\tau}), h_i(S)\le 0, i=1,..,m
\end{align*}
can be converted into an optimisation
\begin{align*}
\max_S \tilde{f}(\log S)
\text{ s.t. } \log S \in \log R(\bar{\tau}), \tilde{h}_i(\log S)\le 0, i=1,..,m
\end{align*}
where $\tilde{f}(z) = f(\exp(z))$ (so, in particular, $\tilde{f}(\log S)=f(S)$), $\log S(\T) = [\log s_1\ ...\ \log s_n]^T$, $\log R = \{\log s: s\in R\}$ and $\tilde{h}_i(z)=h(\exp(z))$.   Provided $-\tilde{f}(\cdot)$ and the $\tilde{h}_i(\cdot)$ are convex functions, the optimisation is a convex problem to which standard tools can then be applied. From this point of view it now follows that we can naturally extend the congestion and contention control ideas of \cite{calderbank06} to the more general scenario considered in \cite{David_TON_2007,clifford06}.

In particular, for the standard family of utility fairness functions given for $w>0$, $\alpha \ge 1$ and $z>0$ by
\begin{align*}
f_{w,\alpha}(z)=
\begin{cases}
w z^{1-\alpha}/(1-\alpha) & \text{if } \alpha\neq 1, \\
w \log(z) & \text{if } \alpha=1,
\end{cases}
\end{align*}
we have $\tilde{f}_{w,\alpha}(z)=f_{w,\alpha}(\exp(z))$ is concave for all $\alpha \geq 1$. In the $\alpha > 1$ case we also get strict concavity of $f$, and the existence and uniqueness of utility fair solutions immediately follows from our log-convexity result. For $\bar{\tau}=1$ an analysis of the boundary of the log rate-region also allows one to show uniqueness of the solution in the case of $\alpha=1$.

\section{Conclusions}
In this paper we establish the log-convexity of the rate region in 802.11 WLANs.   This generalises previous results for Aloha networks and has immediate implications for optimisation based approaches to the analysis and design of fair throughput allocation schemes in 802.11 wireless networks.

{}


\begin{thebibliography}{}

\bibitem{sasha06}
P.~Gupta, A.~L.~Stolyar, ``Optimal Throughput Allocation in General Random-Access Networks,'' 
\emph{Proc. CISS}, 2006.

\bibitem{tassiulas04}
K.~Kar, S.~Sarkar, L.~Tassiulas, ``Achieving Proportional Fairness Using Local Information in Aloha Networks,''  \emph{IEEE Trans. Auto. Control}, 49(10), pp. 1858--1862, 2004.

\bibitem{calderbank06}
J.~W.~Lee, M.~Chiang, A.~R.~Calderbank, ``Jointly Optimal Congestion and Contention Control Based on Network Utility Maximimization,''  \emph{IEEE Comm. Letters}, 10(3), pp. 216--218, 2006.

\bibitem{David_TON_2007}
 D.~Malone, K.~Duffy, and D.~Leith, ``Modeling the 802.11 Distributed Coordination
 Function in Nonsaturated Heterogeneous Conditions," \emph{IEEE/ACM Trans. Networking}, 15(1), pp. 159--172, 2007.
 
 \bibitem{clifford06}
P.~Clifford, K.~Duffy, J.~Foy, D.~J. Leith, and D.~Malone, ``Modeling
 802.11e for data traffic parameter design," \emph{Proc. RAWNET}, 2006.


\bibitem{siris06}
V.~A.~Siris, G.~Stamatakis, ``Optimal CWmin Selection for Achieving Proportional Fairness in Multi-Rate 802.11e WLANs,'' \emph{Proc.  WinTECH}, 2006.

\bibitem{wang04}
X.~Wang, K.~Kar, ``Distributed Algorithms for Max-Min Fair Rate Allocation in ALOHA Networks,''  \emph{Proc. Allerton Conference}, 2004.

\bibitem{wang06}
X.~Wang, K.~Kar, J.~S.~Pang, ``Lexicographic Max-Min Fair Rate Allocation in Random Access Wireless Networks,''  \emph{Proc. IEEE CDC}, 2006.

\bibitem{boudec07}
B.~Radunovic, J.-Y.~Le~Boudec, ``A unified framework for max-min and min-max fairness with applications," 
\emph{IEEE/ACM Trans. Networking}, 15(5), pp. 1073--1083, 2007.

\end{thebibliography}
 \end{document}